\documentclass[twocolumn,a4paper,preprintnumbers,amsmath,amssymb,nofootinbib,floatfix]{revtex4}

\usepackage{graphicx}
\usepackage{dcolumn}
\usepackage{bm}
\usepackage{latexsym}
\usepackage{epsfig}
\usepackage{pbox}
\usepackage{multirow}
\usepackage{rotating}
\usepackage{bm}

\newcommand{\be}{\begin{equation}}
\newcommand{\ee}{\end{equation}}
\newcommand{\beqq}{\setlength\arraycolsep{2pt}\begin{eqnarray}}
\newcommand{\eeqq}{\vspace{0cm} \end{eqnarray}}
\newcommand{\bea}{\begin{eqnarray}}
\newcommand{\eea}{\end{eqnarray}}

\begin{document}

\title{Casimir effect for Elko fields}

\author{S. H. Pereira} \email{shpereira@gmail.com}
\author{J. M. Hoff da Silva} \email{hoff@feg.unesp.br}
\author{Rubia dos Santos} 
\affiliation{Universidade Estadual Paulista (Unesp)\\Faculdade de Engenharia, Guaratinguet\'a \\ Departamento de F\'isica e Qu\'imica\\ Av. Dr. Ariberto Pereira da Cunha 333\\
12516-410 -- Guaratinguet\'a, SP, Brazil}




\begin{abstract}
The Casimir effect for {mass dimension one fermion fields (sometimes called Elko)} in $3+1$ dimension is obtained using Dirichlet boundary conditions. It is shown the existence of a repulsive force four times greater than the case of the scalar field. The precise reason for such differences are highlighted and interpreted, as well as the right parallel of the Casimir effect due to scalar and fermionic fields. 
\end{abstract}

\maketitle

\section{Introduction}
In 1948 the dutch physicist Hendrik Brugt Gerhard Casimir studied a fully quantum effect whose macroscopic manifestation turned out to be one of the most interesting and direct consequences of the existence of quantum vacuum energy \cite{casimir}. The so-called Casimir Effect consists in the appearing of a force between, for instance, two parallel, neutral and perfectly conducting plates placed in vacuum (see \cite{plunien,bordag} for good reviews). Such attractive force appears due to the zero point energy difference for the quantum fields when calculated with external boundary conditions and in the absence of the boundary. Both energies are infinite when calculated from the quantum field theory techniques, but its difference is finite after some renormalization procedure, giving rise to the effect. The theoretical model developed by Casimir was first tested in laboratory by Sparnaay \cite{sparnaay} ten years latter. Since then the effect has been tested by several new experiments, different geometries or boundaries, and also for different fields, as scalars, fermionic and electromagnetic, in both massive and non-massive cases (see \cite{farina,seye,milton} for additional references). 

The Casimir effect brings several interesting properties. It is well known that the sign of the Casimir force is strongly dependent on the geometry of the boundary, on the kind of field and also on the number of spatial dimensions \cite{milton}. For $(3+1)$ dimensions the Casimir force is
strictly attractive between parallel plates, whether one is dealing with conductors or dielectrics, for fields endowed with spin 0 or 1. Different boundary conditions  {in this case}, as Dirichlet and Neumann, also leads to same results. The Casimir force inside a spherical shell was shown to be repulsive by Boyer \cite{boyer}. In Kaluza-Klein theories in $(4 + N)$ dimensions, with $N$ compactified extra dimensions, the force is attractive for $N=1$, while for higher odd dimensions (up to $19$) the Casimir force is repulsive. For $N>21$ the force becomes increasingly attractive. For even $N$ extra dimensions the result is divergent \cite{milton}. In the case of fermionic Dirac fields the results are also notable. Since the Dirac equation is a first order differential equation, the usual boundary conditions are impossible to be used. A much more convenient boundary condition is the so-called MIT Bag Model boundary condition, which implies that there is no flux of fermions through the boundary. The Casimir energy for a massive Dirac fermionic field was computed by Mamayev and Trunov in 1980 \cite{mamayev}, showing to be  {negative, which corresponds to an attractive force.}

\begin{table*}[ht]
{\begin{tabular}{@{}ccccc@{}} \toprule
Field & Force & Energy sign & BC & C/AC relation\\
\colrule
Scalar\hphantom{00} & \hphantom{0}attractive & \hphantom{0}negative & Dirichlet & C\\
Dirac Fermion\hphantom{00} & \hphantom{0}attractive & \hphantom{0}negative & \hphantom{0}Bag Model & AC\\
Elko\hphantom{0} & \hphantom{0}repulsive & \hphantom{0}positive & \hphantom{0}Dirichlet & AC \\\botrule
\end{tabular}\label{ta1} }
\caption{Comparison of Casimir force and energy for different fields, boundary conditions (BC) and commutation (C) or anti-commutation (AC) relation. We refer to parallel plates in $3+1$ dimensions.}
\end{table*}

All these peculiarities show that the study of the Casimir effect is, indeed, an interesting issue concerning the quantum characteristics of the corresponding field. It is important to emphasize, for our future reference, that the comparison of the effect strength between different fields must be taken with some care. For instance, the comparison between  Casimir effect for Dirac fermions and scalar fields must take into account the different boundary conditions for the calculation. 

In this letter we present the calculation of the Casimir effect for the {class of mass dimension one fermion fields discovered more or less recently by Ahluwalia and Grumiller \cite{AHL1,AHL2,AHL3}. We will refer to this new fermionic field just as Elko field, from \textit{Eigenspinoren des ladungskonjugationsoperators}}, whose mass dimension one and neutrality (in which concern to gauge charges) makes it a quantum field candidate to describe dark matter. There are many interesting properties concerning these spinorial objects and associated quantum field (for an interesting and up to date work, including a redefinition of the adjoint leading to another well behaved field, see \cite{AHL4}). For our purposes here, we shall highlight the fact that although Elko fields are objects of spin $1/2$ by construction, its dynamics is dictated by the Klein-Gordon equation only. Therefore, it is possible to use the same boundary condition as the scalar field. Nevertheless, being a spinor, it satisfies a fermionic anti-commutation relation, making the study of its Casimir effect a very interesting question. These half-to-half, in a manner of speaking, characteristics of Elko fields are eminently present in the Casimir effect associated to it. In fact, by studying the simplest case of a non-interacting Elko field of mass $m$ inside a cavity of volume $L^2d$ (assuming $L\gg d$), and imposing Dirichlet boundary conditions on the surfaces, we arrive at the same problem of a massive scalar field in a similar condition. Notice that the analogy between these cases may be insightful, but is restricted to the calculations. There are two crucial departures in the output, namely: the strength of the effect is four times the usual, a result whose cause rests upon the fact that the expansion coefficients of the field are, indeed, a complete set of eingenspinors of the charge conjugation operator. The Casimir energy is positive and related force is repulsive (the opposite of the scalar field case). Due to the very fermionic character of the field we must use anti-commuting relation, contrary to commuting one for scalar fields. Since the boundary condition is now Dirichlet (for a Dirac fermion it is a Bag Model boundary condition), a minus sign survives up the end of calculations. The situation concerning the Casimir energy and force according to the field at hand may be summarized in Table I.

\section{The Elko field and its Casimir energy}
We start by considering the Lagrangian density for a free Elko quantum field ${\eta}(x)$ and its adjoint  $\stackrel{\neg}{\eta}(x)$ (we use natural system of units in which $\hbar=c=1$):
\begin{eqnarray}
\mathcal{L} = \partial^{\mu} \stackrel{\neg}{\eta}(x) \partial_{\mu} \eta(x) - m^2 \stackrel{\neg}{\eta}(x) \eta(x)\,,\label{lagrange}
\end{eqnarray}
with the corresponding Klein-Gordon equation of motion
\begin{equation}
\big(\partial^\mu\partial_\mu + {m^2}\big)\eta(x)=0\,,\label{eqmotion}
\end{equation}
and a similar one for $\stackrel{\neg}{\eta}(x)$.

The quantum fields can be expanded as:
\begin{eqnarray}
\eta(x) &=& \frac{1}{{(2\pi)^{3/2}}} \int\frac{d^3k}{\sqrt{2m\omega_k}}\sum_{ \beta} [ a_{ \beta}({\bf k}) \lambda_{\beta}^{S}({\bf k}) e^{-i k \cdot x} \nonumber\\
&+& a_{\beta}^{\dagger}({\bf k}) \lambda_{\beta}^{A}({\bf k}) e^{i k\cdot x} ]\,,\\
\stackrel{\neg}{\eta}(x) &=& \frac{1}{{(2\pi)^{3/2}}} \int \frac{d^3k}{\sqrt{2m\omega_k}}\sum_{ \beta} [ a_{ \beta}^{\dagger}({\bf k}) \stackrel{\neg}{\lambda}_{\beta}^{S}({\bf k}) e^{i k\cdot x}\nonumber\\&+& a_{\beta}({\bf k}) \stackrel{\neg}{\lambda}_{\beta}^{A}({\bf k}) e^{-i k\cdot x} ]\,,
\end{eqnarray}
where $\omega_k=\sqrt{{\bf k}^2 +m^2} $ and $\lambda_{\beta}^{S/A}({\bf k})$ are the Elko fields satisfying $C\lambda_{\beta}^{S/A}({\bf k})=\pm \lambda_{\beta}^{S/A}({\bf k})$, with $C$ the charge conjugation operator. The positive sign stands for the self-conjugate spinor $(S)$ whilst the negative sign stands for its anti self-conjugate counterpart $(A)$. Throughout the work $\beta=(\{+,-\}, \{-,+\})$ denotes the helicity. Better said, the complete set of Elko fields comprises four spinors: two of them corresponding to different $\beta$ choice for the $S$ case, and two for the $A$ case. The spinor and its dual satisfy (actually, the dual is constructed to satisfy \cite{AHL3,AHL4})
\begin{eqnarray}
\stackrel{\neg}{\lambda}_{\beta'}^{S}({\bf k})  \lambda_{\beta}^{S}({\bf k}) &=& 2 m \delta_{\beta \beta'}\,, \nonumber\\
\stackrel{\neg}{\lambda}_{\beta'}^{A}({\bf k})  \lambda_{\beta}^{S}({\bf k}) &=& -2 m \delta_{\beta \beta'}\,. \label{preserve}
\end{eqnarray}

In order to achieve a positive-definite Hamiltonian associated to the free field, the operators $a_{ \beta}({\bf k})$ are shown to satisfy the usual anti-commutation relations for fermions:
\begin{eqnarray}
\{ a_{ \beta}({\bf k})\,,a_{ \beta'}^\dagger({\bf k}')\}&=& \delta_{\beta \beta'}\delta({\bf k}-{\bf k}')\,,\nonumber \\
\{ a_{ \beta}({\bf k})\,,a_{ \beta'}({\bf k}')\}&=&\{ a_{ \beta}^\dagger({\bf k})\,,a_{ \beta'}^\dagger({\bf k}')\}=0\,.\label{anticomut}
\end{eqnarray}
The operator $a_{ \beta}({\bf k})$ satisfies  $a_{ \beta}({\bf k})\vert 0\rangle=0$, where $\vert 0\rangle$ is the vacuum state without boundary conditions.

The energy-momentum tensor for the field is obtained as usual:
\begin{eqnarray}
T_{\nu}^{\mu} =  \frac{\partial \mathcal{L} }{\partial (\partial_{\mu} \eta)} \partial_{\nu} \eta + \partial_{\nu} \stackrel{\neg}{\eta} \frac{\partial \mathcal{L} }{\partial (\partial_{\mu} \stackrel{\neg}{\eta})} - \delta_{\nu}^{\mu} \mathcal{L},
\end{eqnarray}
for which the zero point energy of the field in free space (without any boundary condition) reads straightforwardly 
\begin{eqnarray}
E_0&=&\int d^3x \langle 0\vert T_{00} \vert 0\rangle\nonumber \\
&=&-4 \frac{L^2d}{(2\pi)^3}\int{d^2k_\Vert} \int_{-\infty}^\infty {dk_z}\, \frac{1}{2}\sqrt{k_\Vert^2+k_z^2+m^2}\,,\nonumber \\\label{energyk}
\end{eqnarray}
where $k_\Vert$ stands for the momentum components parallel to the plates, namely $k_\Vert^2 = k_x^2+k_y^2$. As one can see from the general form of Eq. (\ref{energyk}), the zero point energy already present the $-4$ factor in front of the zero point energy associated to a scalar field. As mentioned, tracing back its origin one sees that the negative sign is a genuine property of the fermionic character of Elko fields, while the four Elko fields defined by means of its neutrality give rise to the aforementioned multiplicity. 

Now, in order to consider the field inside two parallel plates separated by a distance $d$, we impose that the field obeys a boundary condition as:
\begin{equation}
\eta(t,0)=\eta(t,d)=0\,,\label{boundary}
\end{equation}
and the same condition for $\stackrel{\neg}{\eta}$. The quantum fields that satisfies such conditions are then
\begin{eqnarray}
\eta(x) &=&  \int \frac{d^2k_\Vert}{2\pi}\sum_{n=1}^\infty\frac{1}{\sqrt{2m\omega_n d}}\sum_{ \beta} [ a_{ \beta,n} \lambda_{\beta}^{S}({\bf k}) e^{-i \omega_n t} \nonumber\\
&&+ a_{\beta,n}^{\dagger} \lambda_{\beta}^{A}({\bf k}) e^{i \omega_n t} ]\sin(k_n x)\,,\\
\stackrel{\neg}{\eta}(x) &=&  \int \frac{d^2k_\Vert}{2\pi}\sum_{n=1}^\infty\frac{1}{\sqrt{2m\omega_n d}}\sum_{ \beta} [ a_{ \beta,n}^{\dagger} \stackrel{\neg}{\lambda}_{\beta}^{S}({\bf k}) e^{i \omega_n t} \nonumber\\
&&+ a_{\beta,n} \stackrel{\neg}{\lambda}_{\beta}^{A}({\bf k})e^{-i \omega_n t} ]\sin(k_n x)\,,
\end{eqnarray}
provided $\omega_n\equiv \sqrt{k_\Vert^2+k_n^2+m^2}$, with $k_n=n\pi /d$, $n=1,2, \cdots$ and $a_{ \beta,n}\equiv a_{ \beta}({ k}_\Vert,k_n)$. The anti-commutation relation (\ref{anticomut}) must be replaced by
\begin{eqnarray}
\{ a_{ \beta,n}\,,a_{ \beta',n'}^\dagger \}&=&\delta_{\beta \beta'}\delta_{nn'}\delta({ k}_\Vert-{ k}_\Vert')\,,\label{anticomutn}
\end{eqnarray}
where the operator $a_{ \beta,n}$ satisfies  $a_{ \beta,n}\vert 0_B\rangle=0$, where $\vert 0_B\rangle$ is the vacuum state with boundary conditions.

The zero point energy of the field with boundary conditions (\ref{boundary}) is
\begin{eqnarray}
E_{0B}&=&L^2\int_{0}^{d}dx \langle 0_B\vert T_{00} \vert 0_B\rangle\nonumber \\
&=&  -4 \frac{L^2}{(2\pi)^2}\int{d^2k_\Vert}\sum_{n=0}^\infty\, \frac{1}{2}\sqrt{k_\Vert^2+k_n^2+m^2}\,.\nonumber \\\label{energyn}
\end{eqnarray}
It is easy to see that both energies (\ref{energyk}) and (\ref{energyn}) are infinite and thus, some renormalization procedure must be applied to remove the divergences. This procedure can be implemented by dimensional regularization \cite{Amb}, or the simply adoption of a well behaved cutoff. 

The Casimir energy per unit surface is related to the difference between the energies (\ref{energyn}) and (\ref{energyk}):
\begin{eqnarray}
E_{0}^{(cas)} &=&\frac{E_{0B} - E_{0}}{L^2}\nonumber \\
&=&\frac{-2}{(2\pi)^2}\int d^2k_\Vert \Bigg\{\sum_{n=0}^\infty\,\sqrt{k_\Vert^2+(n\pi/d)^2+m^2}\nonumber\\
&&-\int_0^\infty dn\sqrt{k_\Vert^2+(n\pi/d)^2+m^2}\Bigg\}\,\label{eq14}
\end{eqnarray}
where we have used $k_z=n\pi/d$ and $dn=(d/\pi)dk_z$ in (\ref{energyk}).
{Notice that the $n=0$ term does not depends on the separation $d$, not contributing to the force.}
From now on all the calculations strictly falls into that one of a standard scalar field \cite{plunien}. In order to integrate the above expression we make a change of variable $y=k_\Vert^2d^2/\pi^2+n^2+\mu^2$ with $\mu=md/\pi$ and introduce a cutoff function $\exp(-\lambda\sqrt{y})$ depending on the $\lambda=\pi/dk_c$ parameter that must be taken null at the end of calculations. Here, $k_c$ denotes a cutoff wave number. The expression (\ref{eq14}) can be rewritten as $E_{0}^{(cas)}=\lim_{\lambda\to 0}E_0(d,\mu,\lambda)$, with:
\begin{equation}
E_0(d,\mu,\lambda)=-\frac{L^2\pi^2}{8d^3}\Bigg\{ \sum_{n=1}^\infty b(n,\mu,\lambda)-\int_0^\infty dn\, b(n,\mu,\lambda) \Bigg\}\label{eq15}
\end{equation}
and
\begin{equation}
b(n,\mu,\lambda)=\int_{\mu^2+n^2}^\infty dy \sqrt{y}\,\textrm{e}^{-\lambda \sqrt{y}}\,.
\end{equation}
{Now we take advantage of the Poisson's sum formula for a symmetric function\cite{plunien}
\begin{eqnarray}
\sum_{n=1}^\infty b(n,\mu,\lambda)&=&-{1\over 2}b(0,\mu,\lambda)+ \int_0^\infty dn\, b(n,\mu,\lambda) \nonumber\\
&&+ 2\pi \sum_{n=1}^\infty {1\over \pi}\int_0^\infty dn \cos(2\pi n^2)b(n,\mu,\lambda)\,\nonumber\\.\label{poi}
\end{eqnarray}
The first term on the right-hand side of Eq. (\ref{poi}) is $d$ independent while the second one cancels exactly the integral appearing in Eq. (\ref{eq15}). The remaining integral can be performed\cite{plunien}
}
and we obtain the cutoff-independent Casimir energy per unit surface for the Elko field {given by}
\begin{equation}
E_{0}^{(cas)}(\lambda\to 0)=+\frac{m^2}{2\pi^2 d}\sum_{n=1}^\infty\frac{1}{n^2}K_2(2dmn)\,,
\end{equation}
where $K_2(w)$ is the modified Bessel function. For completeness, we emphasize that in the limit $m\ll d^{-1}$ we have $K_2(2mdn)=2(mdn)^{-2}-1/2+O(m^2)$, so that after $n$-summation we have 
\begin{equation}
E_{0}^{(cas)}(\lambda\to 0)=+\frac{\pi^2}{360 d^3}-\frac{m^2}{24 d}+\dots\,.
\end{equation}
On the other hand, in the limit $m\gg d^{-1}$ the Casimir energy is found to be exponentially small
\begin{equation}
E_{0}^{(cas)}(\lambda\to 0)=\frac{m^2}{4\pi^2 d}\bigg(\frac{\pi}{md}\bigg)^{1/2}\textrm{e}^{-2md}\,.
\end{equation}

\section{Conclusion}
As already remarked, the fermionic character of Elko fields along with the fact that it obeys only Klein-Gordon dynamical equation, gives a half-to-half behavior for this particular field. In this application to the study of Casimir effect, it is quite evident such a behaviour by tracing back the reason of the $-4$ term in front of the zero point energies in equations (8) and (13). Such change on the sign of the energy (and consequently to the force) compared to the scalar field is due to the use of anti-commuting relations for the field. 

We shall finalize by pointing out that in Ref. \cite{AHL4} a new dual is proposed in the spinorial formal structure of the field, leading to important consequences in the corresponding quantum field. Some additional mathematical support to the adopted procedure was given in \cite{RJ} and further impact in quantum field theory was analyzed in \cite{WT}. The relevant fact to our study is that such a new dual is introduced in such a way that the relations (\ref{preserve}) are preserved. Therefore, it is indeed expected that all the results founded here can be directly transposed to the fields described in Ref. \cite{AHL4}.


\begin{acknowledgements}
 SHP is grateful to CNPq - Conselho Nacional de Desenvolvimento Cient\'ifico e Tecnol\'ogico, Brazilian research agency, for financial support, grants numbers 304297/2015-1 and 400924/2016-1. JMHS thanks to CNPq for financial support, grants number 445385/2014-6 and 304629/2015-4. RdS thanks to CNPq (34194). 
\end{acknowledgements}


\end{document}